\documentclass[twocolumn]{aastex61}
\pdfoutput=1 
\usepackage{amsmath,amstext}
\usepackage[T1]{fontenc}
\usepackage{apjfonts} 
\usepackage[figure,figure*]{hypcap}
\usepackage{multirow}
\usepackage{booktabs}
\usepackage{hyperref}
\usepackage{cleveref}

\newcommand{\chandra}{\textit{Chandra X-ray Observatory}}
\newcommand{\Chandra}{\textit{Chandra}}
\newcommand{\fermi}{\textit{Fermi}}
\newcommand{\integral}{\textit{INTEGRAL}}
\newcommand{\swift}{\textit{Swift}}

\def\thetav{\theta_{\rm obs}}
\def\thetan{\theta_{1/2,\,\rm jet}}

\def\ep{E_{\rm peak}}
\def\epobs{E_{\rm peak,\,off}}

\def\davg{\langle D \rangle}

\newcommand{\package}[1]{\textsc{#1}}

\shorttitle{ALMA and GMRT observations of GW170817}
\shortauthors{Kim et al.}

\begin{document}

\title{ALMA and GMRT constraints on the off-axis gamma-ray burst 170817A from the binary neutron star merger GW170817}

\author{S. Kim}
\affiliation{Instituto de Astrof{\'{\i}}sica and Centro de Astroingenier{\'{\i}}a, Facultad de F{\'{i}}sica, Pontificia Universidad Cat{\'{o}}lica de Chile, Casilla 306, Santiago 22, Chile}
\affiliation{Max-Planck-Institut f\"{u}r Astronomie K\"{o}nigstuhl 17 D-69117 Heidelberg, Germany}

\author{S. Schulze}
\affiliation{Department of Particle Physics and Astrophysics, Weizmann Institute of Science, Rehovot 761000, Israel}

\author{L. Resmi}
\affiliation{Indian Institute of Space Science \& Technology, Trivandrum 695547, India}

\author{J. Gonz{\'{a}}lez-L{\'{o}}pez}
\affiliation{Instituto de Astrof{\'{\i}}sica and Centro de Astroingenier{\'{\i}}a, Facultad de F{\'{i}}sica, Pontificia Universidad Cat{\'{o}}lica de Chile, Casilla 306, Santiago 22, Chile}

\author{A.~B.~Higgins}
\affiliation{Department of Physics and Astronomy, University of Leicester, University Road, Leicester, LE1 7RH, United Kingdom}

\author{C.~H. Ishwara-Chandra}
\affiliation{National Center for Radio Astrophysics, Pune 411007, India}

\author{F. E. Bauer}
\affiliation{Instituto de Astrof{\'{\i}}sica and Centro de Astroingenier{\'{\i}}a, Facultad de F{\'{i}}sica, Pontificia Universidad Cat{\'{o}}lica de Chile, Casilla 306, Santiago 22, Chile}
\affiliation{Millennium Institute of Astrophysics (MAS), Nuncio Monse{\~{n}}or S{\'{o}}tero Sanz 100, Providencia, Santiago, Chile}
\affiliation{Space Science Institute, 4750 Walnut Street, Suite 205, Boulder, Colorado 80301, USA}

\author{I. de Gregorio-Monsalvo}
\affiliation{European Southern Observatory, Alonso de C\'ordova 3107, Vitacura, Santiago 763-0355, Chile}
\affiliation{Joint ALMA Observatory, Alonso de C\'ordova 3107, Vitacura, Santiago 763-0355, Chile}

\author{M. De Pasquale}
\affiliation{Department of Astronomy and Space Sciences, Istanbul University, 34119 Beyaz\i t, Istanbul, Turkey} 

\author{A. de Ugarte Postigo}
\affiliation{Instituto de Astrof\'isica de Andaluc\'ia (IAA-CSIC), Glorieta de la Astronom\'ia, s/n, 18008, Granada, Spain}

\author{D. A. Kann}
\affiliation{Instituto de Astrof\'isica de Andaluc\'ia (IAA-CSIC), Glorieta de la Astronom\'ia, s/n, 18008, Granada, Spain}

\author{S. Mart\'in}
\affiliation{European Southern Observatory, Alonso de C\'ordova 3107, Vitacura, Santiago 763-0355, Chile}
\affiliation{Joint ALMA Observatory, Alonso de C\'ordova 3107, Vitacura, Santiago 763-0355, Chile}

\author{S. R. Oates}
\affiliation{Department of Physics, University of Warwick, Coventry, CV4 7AL, United Kingdom}

\author{R. L. C. Starling}
\affiliation{Department of Physics and Astronomy, University of Leicester, University Road, Leicester, LE1 7RH, United Kingdom}

\author{N. R. Tanvir}
\affiliation{Department of Physics and Astronomy, University of Leicester, University Road, Leicester, LE1 7RH, United Kingdom}

\author{J. Buchner}
\affiliation{Millennium Institute of Astrophysics (MAS), Nuncio Monse{\~{n}}or S{\'{o}}tero Sanz 100, Providencia, Santiago, Chile}
\affiliation{Instituto de Astrof{\'{\i}}sica and Centro de Astroingenier{\'{\i}}a, Facultad de F{\'{i}}sica, Pontificia Universidad Cat{\'{o}}lica de Chile, Casilla 306, Santiago 22, Chile}

\author{S. Campana}
\affiliation{INAF - Osservatorio astronomico di Brera, Via E. Bianchi 46, 23807, Merate (LC), Italy}

\author{Z. Cano}
\affiliation{Instituto de Astrof\'isica de Andaluc\'ia (IAA-CSIC), Glorieta de la Astronom\'ia, s/n, 18008, Granada, Spain}

\author{S. Covino}
\affiliation{INAF - Osservatorio astronomico di Brera, Via E. Bianchi 46, 23807, Merate (LC), Italy}

\author{A. S. Fruchter}
\affiliation{Space Telescope Science Institute, 3700 San Martin Dr., Baltimore, MD 21218}

\author{J. P. U. Fynbo}
\affiliation{Dark Cosmology Centre, Niels Bohr Institute, University of Copenhagen, Juliane Maries Vej 30, DK-2100 Copenhagen \O, Denmark}

\author{D. H. Hartmann}
\affiliation{Department of Physics and Astronomy, Clemson University, Clemson, SC 29634-0978, USA}

\author{J. Hjorth}
\affiliation{Dark Cosmology Centre, Niels Bohr Institute, University of Copenhagen, Juliane Maries Vej 30, DK-2100 Copenhagen \O, Denmark}

\author{P. Jakobsson}
\affiliation{Centre for Astrophysics and Cosmology, Science Institute, University of Iceland, Dunhagi 5, 
107 Reykjav\'ik, Iceland}

\author{A. J. Levan}
\affiliation{Department of Physics, University of Warwick, Coventry, CV4 7AL, United Kingdom}

\author{D. Malesani}
\affiliation{Dark Cosmology Centre, Niels Bohr Institute, University of Copenhagen, Juliane Maries Vej 30, DK-2100 Copenhagen \O, Denmark}

\author{M. J. Micha{\l}owski}
\affiliation{Astronomical Observatory Institute, Faculty of Physics, Adam Mickiewicz University, ul.~S{\l}oneczna 36, 60-286 Pozna{\'n}, Poland}

\author{B. Milvang-Jensen}
\affiliation{Dark Cosmology Centre, Niels Bohr Institute, University of Copenhagen, Juliane Maries Vej 30, DK-2100 Copenhagen \O, Denmark}

\author{K. Misra}
\affiliation{Aryabhatta Research Institute of observational sciencES (ARIES), Manora Peak, Nainital 263 001, India}

\author{P. T. O'Brien}
\affiliation{Department of Physics and Astronomy, University of Leicester, University Road, Leicester, LE1 7RH, United Kingdom}

\author{R. S\'anchez-Ram\'irez}
\affiliation{INAF, Istituto Astrofisica e Planetologia Spaziali, Via Fosso del Cavaliere 100, I-00133 Roma, Italy}

\author{C. C. Th\"one}
\affiliation{Instituto de Astrof\'isica de Andaluc\'ia (IAA-CSIC), Glorieta de la Astronom\'ia, s/n, 18008, Granada, Spain}

\author{D. J. Watson}
\affiliation{Dark Cosmology Centre, Niels Bohr Institute, University of Copenhagen, Juliane Maries Vej 30, DK-2100 Copenhagen \O, Denmark}

\author{K. Wiersema}
\affiliation{Department of Physics, University of Warwick, Coventry, CV4 7AL, United Kingdom}

\correspondingauthor{Sam Kim, Steve Schulze, Lekshmi Resmi}
\email{skim@astro.puc.cl, steve.schulze@weizmann.ac.il, l.resmi@iist.ac.in}

\begin{abstract}
Binary neutron-star mergers (BNSMs) are among the most readily detectable gravitational-wave (GW) sources with LIGO. They are also thought to produce short $\gamma$-ray bursts (SGRBs), and  kilonovae that are powered by {\it r}-process nuclei. Detecting these phenomena simultaneously would provide an unprecedented view of the physics during and after the merger of two compact objects. Such a Rosetta Stone event was detected by LIGO/Virgo on 17 August 2017 at a distance of $\sim 44$~Mpc. We monitored the position of the BNSM with ALMA at 338.5 GHz and GMRT at 1.4 GHz, from 1.4 to 44 days after the merger. Our observations rule out any afterglow more luminous than $3\times 10^{26}~{\rm erg\,s}^{-1}\,{\rm Hz}^{-1}$ in these bands, probing $>$2--4 dex fainter than previous SGRB limits. We match these limits, in conjunction with public data announcing the appearance of X-ray and radio emission in the weeks after the GW event, to templates of off-axis afterglows. Our broadband modeling suggests that GW170817 was accompanied by a SGRB and that the GRB jet, powered by $E_{\rm AG,\,iso}\sim10^{50}$~erg, had a half-opening angle of $\sim20^\circ$, and was misaligned by $\sim41^\circ$ from our line of sight. The data are also consistent with a more collimated jet: $E_{\rm AG,\,iso}\sim10^{51}$~erg, $\theta_{1/2,\,\rm jet}\sim5^\circ$, $\theta_{\rm obs}\sim17^\circ$. This is the most conclusive detection of an off-axis GRB afterglow and the first associated with a BNSM-GW event to date.  Assuming a uniform top-hat jet, we use the viewing angle estimates to infer the initial bulk Lorentz factor and true energy release of the burst.
\end{abstract}


\section{Introduction}
The existence of gravitational waves (GWs) was predicted  in 1916 \citep{Einstein1916a, Einstein1918a}, but it took almost a century to directly observe them \citep{Abbott2016a}. A type of GW signal readily detectable with the Laser Interferometer Gravitational-wave Observatory (LIGO) is linked to the coalescence of two neutron stars \citep{Abadie2010a}. This class of object is also thought to be the progenitor of short $\gamma$-ray bursts (SGRBs; duration $\lesssim2$~s; \citealt{Eichler1989a, Nakar2007a, Berger2014a}). In addition, the temperatures and densities in the debris of the merger are thought to be high enough to also produce radioactive nuclei through rapid neutron capture. Their decays could give rise to faint supernova-like transients, called kilonovae \citep[KNe; e.g.,][for a review see also \citealt{Metzger2017a}]{Li1998a, Rosswog2005a, Kasen2013a}. Observational evidence for a KN was found in the near-IR photometry of SGRB 130603B \citep{Tanvir2013a} and possibly in optical photometry of GRBs 050709 and 060614 \citep{Jin15,Jin16}. However, without a spectrum, the conjecture that SGRBs are accompanied by KNe and therefore that SGRBs are connected with binary neutron star mergers (BNSMs) is not free of ambiguity. 

On 17 August 2017 at 12:41:04 UTC, the joint LIGO and Virgo observing run detected a BNSM at $40^{+8}_{-14}$ Mpc within an area of 28~deg$^2$ projected on the sky \citep[][]{LIGO_VIRGO_1}.
The precise distance and localization gave the follow-up with optical wide-field imagers a flying start \citep[for a comprehensive review see ][]{MMA2017a}. \citet{Coulter2017a} targeted galaxies at this distance and detected a new object, SSS17a (IAU identification: AT2017gfo; \citealt{Coulter2017b}), at $\alpha_{\rm J2000}=13^{\rm h}09^{\rm m}48\fs09$, $\delta_{\rm J2000}=-23^\circ22'53\farcs3$;  $10\farcs3$ from NGC\,4993 at 43.9~Mpc \citep[$z=0.00984$; ][for a detailed discussion see also \citealt{Hjorth2017a}]{Levan2017a}.\footnote{The luminosity distance was derived for a flat $\Lambda$CDM cosmology with $\Omega_{\rm m} = 0.315$, $\Omega_\Lambda = 0.685$, $H_0 = 67.3~{\rm km~s}^{-1}~\rm{Mpc}^{-1}$ \citep{Planck2014a}. 
We use this cosmology throughout the paper.} This discovery was confirmed by several teams including \citet{Allam2017a}, \citet{Melandri2017a}, \citet{Tanvir2017a} and \citet{Yang2017a}. The transient rapidly faded in the optical, but showed a much slower evolution in the near-IR \citep{Tanvir2017b}. Spectra of AT2017gfo revealed very broad absorption features, due to relativistic expansion velocities \citep{Pian2017a, Smartt2017a, Tanvir2017b}, similar to those expected for KNe \citep{Kasen2013a, Tanaka2014a}. Such features are unlike any known for supernova spectra and strongly argued for a connection between AT2017gfo and GW170817 \citep{Siebert2017a}. 

The GBM detector aboard the $\gamma$-ray satellite \fermi\ \citep{Blackburn2017a, vonKienlin2017a, Goldstein2017a, Goldstein2017b} as well as \integral\ \citep{Savchenko2017a, Savchenko2017b} detected a faint 2-s duration GRB (hereafter GRB 170817A), 1.7~s after GW170817 \citep{GRB2017a}. Although the chance coincidence is very small to find both transients quasi-contemporaneous and in the same region of the sky, the credible region of the $\gamma$-ray localization had a size of $\sim1100$~deg$^2$ (90\% confidence; \citealt{Blackburn2017a}). To firmly establish the connection between GRB 170817A and GW170817 by detecting the afterglow of the GRB in the X-ray and radio bands, numerous groups carried out large follow-up campaigns to very deep limits, but without success \citep[e.g.][]{Alexander2017a, Bannister2017a, Cenko2017a, Corsi2017d, De2017a, Deller2017a, Evans2017a, Kaplan2017a, Margutti2017a, Resmi2017a, Sugita2017a}. Not until nine days after GW170817 a brightening X-ray source was detected at the position of AT2017gfo \citep{Troja2017a}. Subsequent X-ray observations confirmed the brightening \citep{Fong2017a, Haggard2017a}. About a week later, \citet{Corsi2017b} and \citet{Mooley2017a} detected an emerging radio source at 3 and 6~GHz as well. While these observations might support the SGRB connection, such a behavior is highly atypical for GRB afterglows \citep[e.g.,][]{Piran2004a}.

In this letter, we examine the afterglow properties of AT2017gfo. We present sub-mm and radio observations obtained with the Atacama Large Millimeter/submillimeter Array (ALMA) and the Giant Metrewave Radio Telescope (GMRT) between 1.4 and 44.1 days after the GW detection. We augment our dataset with public X-ray, optical, and radio data and confront GRB afterglow models with observations.

All uncertainties reported in this paper are given at $1\sigma$ confidence. Non-detections are reported at $3\sigma$ confidence, unless stated otherwise.

\section{Observations and data reduction}\label{sec:obs}

We observed the field of AT2017gfo as a part of the observing program 2016.1.00862.T (P.I. Kim) with ALMA in the Atacama desert (Chile; \citealt{Wootten2009a}) and as part of the Director's Discretionary Time (DDT) Proposal DDTB285 with the GMRT, Pune (India; \citealt{Swarup1991a}; P.I. Resmi). 

\subsection{ALMA observations}
Our initial ALMA campaign started on 18 August 2017 at 22:50:40 UTC (1.4 days after GW170817) and lasted for eight days \citep{Schulze2017a}. In addition, we secured a final epoch $\sim44$ days after GW170817. In total, we obtained six epochs at 338.5~GHz (Table \ref{tab:data}). The integration time of each observation was set to reach a nominal r.m.s. of $\sim40$~$\mu$Jy/beam. The initial ALMA observations were performed 
in the C40-7 configuration,
with a field of view of $18\farcs34$ in diameter and an average synthesized beam 
of $\approx$$0\farcs13\times0\farcs07$.
The observation at 44 days after GW170817 was performed in the most extended ALMA configuration, C40-8/9, yielding a synthesized beam of $0\farcs026\times0\farcs016$.

The ALMA data were reduced with scripts provided by ALMA and with the software package \package{Common Astronomy Software Applications} (\package{CASA}) version 4.7.2 \citep{McMullin2007a}.\footnote{\href{https://casa.nrao.edu}{https://casa.nrao.edu}} For each epoch, we created images using \package{tclean}, with 
a pixel size of $0\farcs01$/px. We interactively selected cleaning regions around detected sources (none corresponding to the AT2017gfo counterpart). The cleaning process was repeated until no clear emission was left. The r.m.s. was measured in a $10''$-width box around the central position in the images without primary beam correction. 

No significant signal is detected at the position of AT2017gfo. Table \ref{tab:data} summarizes the $3\sigma$ detection limits. The galaxy nucleus is well detected and marginally resolved in our data ($\alpha, \delta\, ({\rm J2000}) = 13^{\rm h}09^{\rm m}47\fs69, -23^\circ23'02\farcs37$). We measure $1.07\pm0.21$~mJy at 338.5~GHz. In addition, we detect a marginally resolved sub-mm galaxy with $F_{338.5\,\rm GHz} = 1.15 \pm 0.21$~mJy at $\alpha, \delta\, ({\rm J2000}) = 13^{\rm h}09^{\rm m}48\fs39, -23^\circ22'48\farcs29$. The quasars J1337$-$1257 and J1427$-$4206 were used for band and flux calibration, and J1256$-$2547, J1937$-$3958, and J1258$-$2219 for phase calibration.

\subsection{GMRT Observations}
The GMRT is one of the most sensitive low-frequency radio telescopes in operation currently.
It operates at low radio frequencies from $150$~MHz to $1.4$~GHz \citep{Swarup1991a}.
We secured three epochs in the L band, centered at 1.39~GHz, between 25 August 2017 and 16 September 2017 (i.e., between 7.9 and 29.8 days since GW170817; Table \ref{tab:data}; \citealt{Resmi2017a}). The observing  time was $\sim1.5$~hr for each observation. The first epoch was performed with the new 200-MHz correlator that divides the bandpass into 2048 channels of which $\sim70\%$ were usable, due to radio frequency interference. The second and the third epoch were performed with the 32-MHz correlator. The synthesized beam sizes were typically $4''\times2''$. The quasar 3C286 was used as flux and bandpass calibrator and J1248$-$199 was used for phase and additional bandpass calibration. Data reduction was carried out with the \package{NRAO Astronomical Image Processing Software}\footnote{\href{http://www.aips.nrao.edu}{http://www.aips.nrao.edu}} (\package{AIPS}; \citealt{Wells1985a}) using standard procedures.

While the marginally resolved nucleus of the host galaxy of AT2017gfo, NGC 4993, was detected with $\sim 570 \mu$~Jy in $1.39$~GHz, no significant signal is detected at the position of AT2017gfo itself. Table \ref{tab:data} summarizes the $3\sigma$ detection limits, where the r.m.s. level is estimated from source-free regions using the task \package{TVSTAT}.

\subsection{Other Observations}

To augment our data set, we incorporated radio measurements from \citet{Corsi2017e, Corsi2017f, Corsi2017b, Corsi2017c, Corsi2017d}, \citet{Hallinan2017a}, \citet{Kaplan2017a}, \citet{Mooley2017a, Mooley2017b} (see also \citealt{Hallinan2017b}), obtained with the Australia Telescope Compact Array (ATCA) and the Very Large Array (VLA). We used the VLA exposure time calculator\footnote{\href{https://obs.vla.nrao.edu/ect/}{https://obs.vla.nrao.edu/ect/}} to convert the relative measurements of \citet{Corsi2017b} and \citet{Mooley2017a} into radio flux densities, adopting r.m.s. values 50\% higher than nominal to mitigate possible losses due to antennae problems and adverse observing conditions. We also included the X-ray constraints of \citet{Evans2017b}, \citet{Haggard2017b} and \citet{Troja:nature} from the {\swift} satellite and {\chandra}, as reported in \citet{MMA2017a}, as well as optical photometry obtained with the \textit{Hubble~Space~Telescope} and ESO's 8.2-m Very Large Telescope (VLT) by \citet{Tanvir2017b}.

\vspace{-0.1in}
\begin{table}
\caption{Log of sub-mm/mm and radio observations of AT2017gfo.}
\vspace{-0.2in}
\begin{center}
\begin{tabular}{ccccc}
\toprule
$T_{\rm start}$	& Epoch 	& Frequency 	& Integration & $F_\nu$\\
(UT) 			& (day)		& (GHz)	  		& time (s) & ($\mu$Jy)\\
\midrule
\multicolumn{5}{c}{\textbf{ALMA}}\\
\midrule
18/08/2017 22:50:40 & 1.4 		& 338.5  		& 	2238		& $<126$ \\ 
20/08/2017 18:19:35 & 3.2 		& 338.5  		& 	2238		& \multirow{2}*{$<90^\dagger$}\\ 
20/08/2017 22:40:16 & 3.4 		& 338.5  		& 	2238		& \\ 
25/08/2017 22:35:17 & 8.4 		& 338.5  		& 	2238		& $<150$\\ 
26/08/2017 22:58:41 & 9.4		& 338.5  		& 	1724		& $<102$\\ 
30/09/2017 15:22:00 & 44.1		& 338.5  		&   1832	    & $<93$\\  
\midrule
\multicolumn{5}{c}{\textbf{GMRT}}\\
\midrule
25/08/2017 09:30:00  & 7.9		& 1.39			& 5400			& $<69$\\
09/09/2017 11:30:00  & 23.0	    & 1.39			& 5400			& $<108$\\
16/09/2017 07:30:00	 & 29.8		& 1.39			& 5400			& $<126$\\
\bottomrule
\end{tabular}
\end{center}
\vspace{-0.15in}
\tablecomments{The epoch is with respect to the time of GW170817.\\ $^\dagger$ The r.m.s. was measured after combining both epochs from 20 August 2017.
}
\label{tab:data}
\vspace{-0.1in}
\end{table}

\section{Results and discussion}
\subsection{GRB 170817A in the context of other SGRBs}\label{sec:31}
\begin{figure}
\includegraphics[width=1\columnwidth]{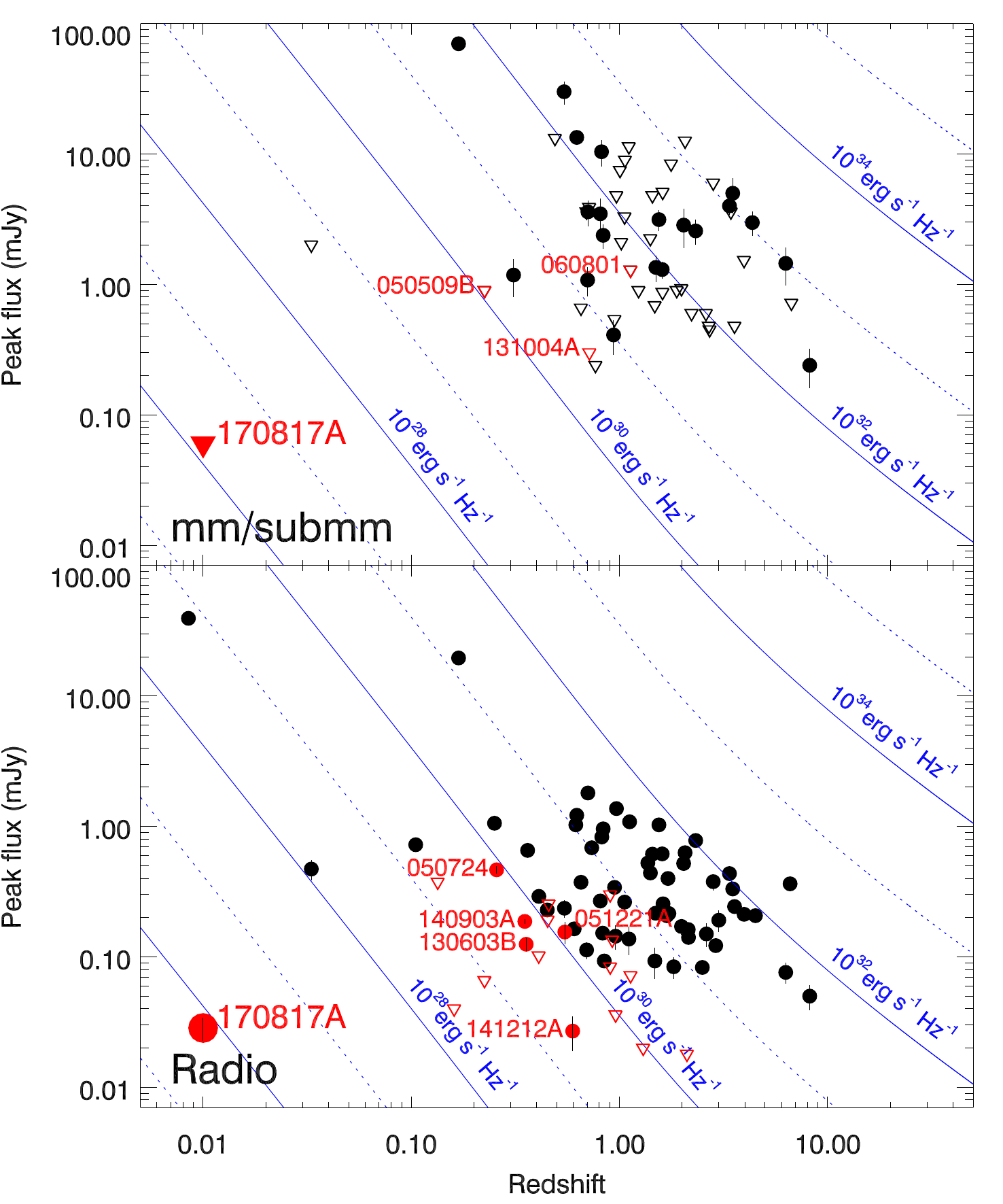}
\caption{
Peak flux densities of GRB afterglows derived from mm/sub-mm (top) and radio (bottom) observations. Filled circles and empty triangles denote detections and 3$\sigma$ limits, respectively. SGRBs are shown in red and LGRBs in black. Solid and dotted blue curves indicate equal afterglow luminosities. The non-detection in the sub-mm corresponds to a luminosity limit of $3\times10^{26}~{\rm erg\,s}^{-1}{\,\rm Hz}^{-1}$. The faint radio counterpart detected by \citet{Corsi2017b} suggests a peak luminosity of $\sim7\times10^{26}~{\rm erg\,s}^{-1}{\,\rm Hz}^{-1}$. }
\label{fig:sub-mm_sample}
\end{figure}

The interaction of the GRB blastwave with the circumburst medium produces an afterglow from X-ray to radio frequencies. The peak of the synchrotron afterglow spectrum is, however, expected to be in the sub-mm/mm band and it rapidly crosses the band towards lower frequencies ($\nu_{\rm m}\propto t^{-3/2}$; \citealt{Sari1998a}). Our initial ALMA 3$\sigma$ limit of $F_{\rm 338.5\,{\rm GHz}}<$126~$\mu$Jy at 1.4~days after GW170817 corresponds to a luminosity of $3\times10^{26}~{\rm erg\,s}^{-1}\,{\rm Hz}^{-1}$ at the redshift of AT2017gfo. Comparing this estimate for the peak flux of GRB 170817A to estimates of other GRBs from \citet{deUgartePostigo2012a} (see Fig. \ref{fig:sub-mm_sample}, top panel), our sub-mm afterglow limits are $\sim$3--4 orders of magnitude fainter than those associated with any long-duration GRBs (LGRBs) or SGRBs.\footnote{We consider all bursts as SGRBs if the burst duration is $<2$~s (observer frame) or if the initial pulse complex lasted less than 2~s (For a critical reflection on the burst duration criterion see \citealt{Bromberg2013a} and for distinguishing between short and long GRBS see \citealt{Zhang2009a} and \citealt{Kann2011a}). In total, 127 \swift\ GRBs, detected until the end of September 2017, fulfilled this criterion. Among those 35 have reliable redshifts:  050509B, 050709, 050724, 051221A, 060502B, 060614, 060801, 061006, 061210, 061217,
070429B, 070714B, 070724, 071227, 080123, 080905A, 090510, 100117A, 100206A, 100625A, 100816A, 101219A,  111117A, 130603B, 131004A, 140622A, 140903A, 141212A, 150120A, 150423A, 150101B, 160410A, 160624A, 160821B, and 170428A.
}

However, the maximum frequency is also correlated with the energy within the jet and the energy release in $\gamma$-rays, e.g., $\nu_m\propto E^{1/2}$ for a constant density circumburst medium \citep[][]{Piran2004a}. \citet{GRB2017a} reported an exceptionally low isotropic-equivalent energy of only $E_{\gamma,\,\rm iso}=(3.08\pm0.72)\times10^{46}$~erg. Hence, it is conceivable that the peak of the afterglow spectrum was already in the radio band during our first ALMA observation. To get an additional estimate of the peak luminosity, we use results in \citet{Corsi2017b}. Their measurement of $F_{6\,{\rm GHz}}\sim28.5~\mu$Jy at 28.5 days after GW170817 translates to a luminosity of $\sim7\times10^{25}~{\rm erg\,s}^{-1}\,{\rm Hz}^{-1}$. Compared to radio measurements of LGRBs and SGRBs in \citet{Fong2015} and \citet{Chandra2012a}, respectively (Fig.~\ref{fig:sub-mm_sample}, bottom panel), the radio afterglow is 2 orders of magnitude fainter than those of LGRBs and SGRBs.

\begin{figure}
\includegraphics[width=1\columnwidth]{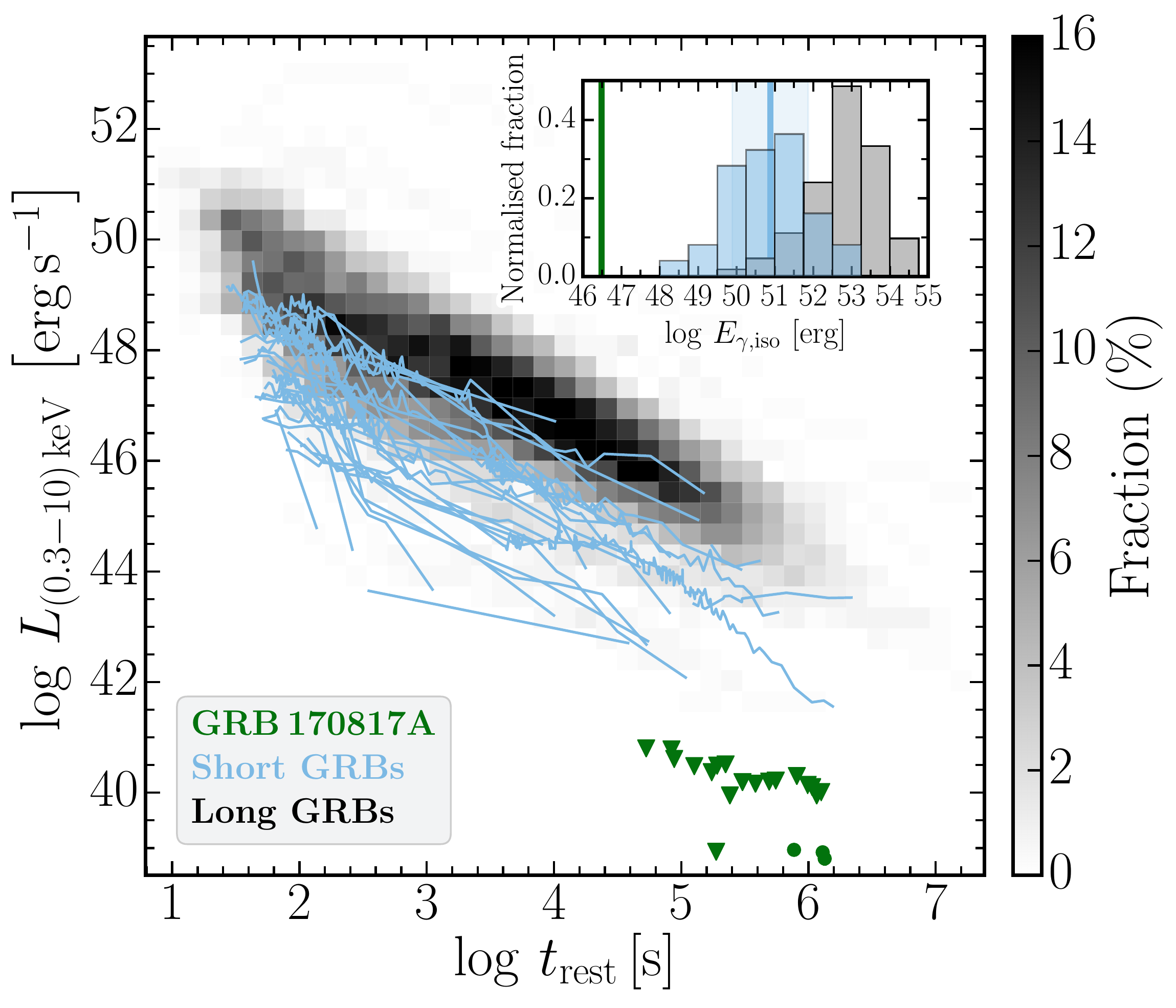}
\caption{
X-ray light curves of GRB afterglows. The parameter space occupied by 402 LGRBs, detected between December 2004 and September 2017, is indicated by the density plot. Light curves of 31 SGRBs with detected X-ray afterglows are shown in light blue. GRB 170817A lies 2.6 orders of magnitude below the other SGRBs. The inset displays the distribution of energy release at $\gamma$-ray energies (SGRBs: blue; LGRBs: gray). The blue vertical line and shaded region indicate the median value and dispersion of the SGRB distribution, respectively.}
\label{fig:Xray_sample}
\end{figure}

While sub-mm and radio observations are direct tracers of the peak of the afterglow spectrum, only $\sim7\%$ of all \swift\ GRBs were bright enough to attempt sub-mm and radio observations. This observational bias is likely to skew the known population towards the bright end of any luminosity function. On the other hand, almost all \swift\ GRBs are detected at X-rays.  The mapping between X-ray brightness and the peak of the afterglow spectrum is more complex. It depends on the location of the cooling break, which is usually between the optical and the X-rays, and the density profile of the circumburst medium. Nonetheless, \citet{deUgartePostigo2012a} showed that X-ray brightness is a useful diagnostic for comparing afterglow luminosities.

To generate an X-ray diagnostic plot, we retrieved the X-ray light curves of 402 LGRBs and 31 SGRBs with detected X-ray afterglows (at least at two epochs) and known redshift from the \swift\ Burst Analyser \citep{Evans2010a}. Identical to \citet{Schulze2014a}, we computed the rest-frame light curves and resampled the light curves of the LGRB sample on a grid (gray shaded region in Fig. \ref{fig:Xray_sample}).  The individual light curves of the SGRB sample are shown in light blue. Already the \swift\ non-detections presented in \citet{Evans2017b} (downward pointing triangles in Fig. \ref{fig:Xray_sample}) revealed that the afterglow is $>1.5$~dex fainter than the faintest SGRB with detected afterglow (downward pointing triangles in Fig. \ref{fig:Xray_sample}). The deep \Chandra\ observation by \citet{Troja:nature} at 2.3~days after GW170817 excluded any afterglow brighter than $>10^{39}~{\rm erg\,s}^{-1}$, i.e., a factor of 10 below the \swift\ upper limits.

To put these limits in context, \citet{Evans2017b} placed 10,000 fake GRBs, generated from the flux-limited SGRB sample in \citet{DAvanzo2014a}, at the distance of GW170817. These authors estimated that the \swift\ X-ray telescope would have detected $\sim65\%$ of all simulations. The deeper \Chandra\ observations probed a larger portion of the parameter space. However, \citet{Rowlinson2010a} showed that a number of SGRBs have extremely rapidly fading X-ray afterglows, which would have evaded detection at the time of the \swift\ and the \Chandra\ observations.

The new quality of the X-ray emission of AT2017gfo is not only its faintness, but actually its emergence more than a week after GW170817. This behavior is inconsistent with known X-ray afterglows. When the X-ray afterglow was detected with \Chandra\ \citep{Troja:nature, Haggard2017b, Margutti2017b}, the luminosity of $L_{(0.3-10)\,\rm keV}\sim8\times10^{38}~{\rm erg\,s}^{-1}$ was still 2.6 dex fainter than that of any GRB \textit{with detected X-ray afterglow}.

This faintness of the afterglow is also reflected in the very low energy release at $\gamma$-rays, $E_{\gamma,\,\rm iso}$. The observed $E_{\gamma,\,\rm iso}$ distribution of SGRBs and LGRBs is shown in the inset of Fig. \ref{fig:Xray_sample}. The vertical blue line and the shaded region display the median $\left[\log\,\left(E_{\rm iso}/{\rm erg}\right) = 50.88\pm0.18\right]$ and the sample dispersion $\left[\sigma\{\log\,\left(E_{\rm iso}/{\rm erg}\right)\} =0.99^{+0.14}_{-0.12}\right]$ of the SGRB sample, computed with \package{(Py)MultiNest} \citep{Feroz2013a, Buchner2014a}. With a prompt energy release of $3.08\times10^{46}$~erg (green vertical line the inset of Fig. \ref{fig:Xray_sample}), GRB 170817A was $\sim1.5$~dex less energetic than the least energetic SGRB known so far and its deviation from the distribution median is $\sim4.4\sigma$.

In conclusion, observations of the afterglow revealed an exceptionally under-luminous afterglow at all wavelengths at the position of AT2017gfo. This extremeness is also reflected in the $\gamma$-ray properties.

\subsection{Modeling the broadband afterglow}\label{sec:modeling}

\begin{figure*}
\includegraphics[width=1\textwidth]{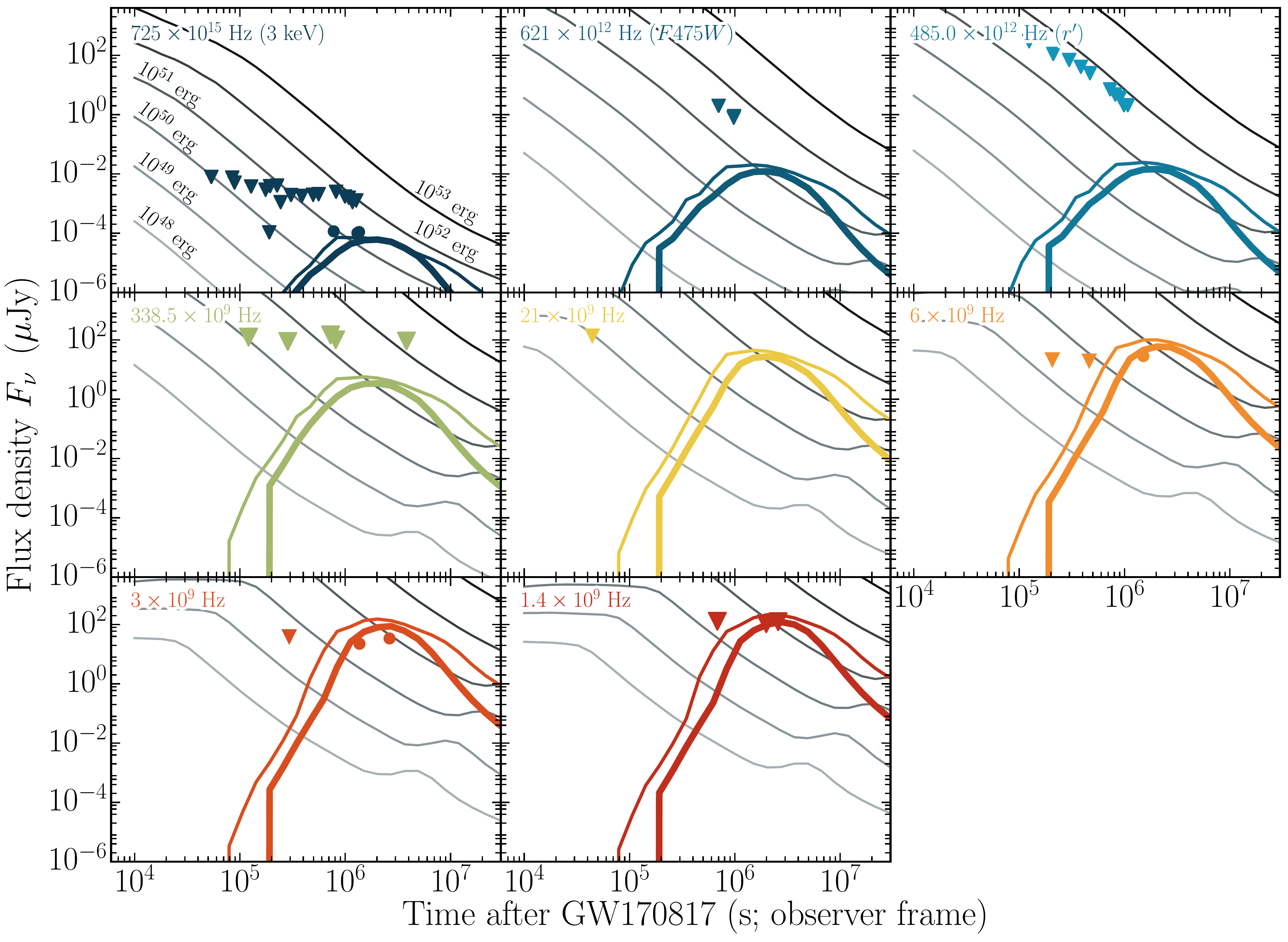}
\caption{The afterglow from radio to X-ray frequencies (detections: $\bullet$; non-detections: $\blacktriangledown$; our ALMA and GMRT measurements are displayed slightly larger.). The light curves are adequately modeled with two distinct templates: \textit{model 1} -- $E_{\rm AG,\, iso}\sim10^{50}$~erg, $\theta_{1/2, \rm jet}\sim20^\circ$, $\theta_{\rm obs}\sim41^\circ$, $n\sim10^{-2}~{\rm cm}^{-3}$, and $\epsilon_B\sim10^{-2}$ (thin curves); \textit{model 2} -- $E_{\rm AG,\, iso}\sim10^{51}$~erg, $\theta_{1/2, \rm jet}\sim5^\circ$, $\theta_{\rm obs}\sim17^\circ$, $n\sim5\times10^{-4}~{\rm cm}^{-3}$, and $\epsilon_B\sim2\times10^{-3}$ (thick curves). The gray curves show the evolution of an on-axis afterglow with $\theta_{1/2, \rm jet}=5^\circ$, $\theta_{\rm obs}=0^\circ$, $n=10^{-2}~{\rm cm}^{-3}$, and $\epsilon_B=10^{-2}$. The energy in the afterglow was varied between $10^{48}$ and $10^{53}$~erg, indicated by the gray-scale color pattern. Combining the non-detections at early times and the detections at late-times rules out the entire parameter space of on-axis afterglow models. The rebrightening seen in some on-axis models at $>10^6$~s is due to the contribution of the GRB counter-jet.
For a detailed discussion see $\S$\ref{sec:modeling}.}
\label{fig:lc}
\end{figure*}

The previous considerations placed the GRB in the context of long- and short-duration GRBs detected by \swift. The discussion neglected the peculiar evolution of the afterglow:
non-detection of the afterglow during the first week and its emergence at later epochs. These properties are highly atypical for GRBs, assuming the GRB jet axis is aligned with our line of sight. In the following, we model the observed evolution from X-rays to radio with templates from two-dimensional relativistic hydrodynamical jet simulations using \package{boxfit} version 2 with the methods described in \citet{vanEerten2012}. The templates are generated from a wide range of physical parameters. Here, we use a nine-parameter model:
\begin{equation}
F_\nu = L\left(E_{\rm AG,\,iso},\,n,\,\theta_{1/2, \rm jet},\,\theta_{\rm obs},\, p,\,\epsilon_e,\,\epsilon_B,\,\xi_N,\,z\right)\nonumber
\end{equation}
where $E_{\rm AG, iso}$ is the isotropic equivalent energy of the blastwave (afterglow)\footnote{In the discussed models, $E_{\rm \, iso}$ always corresponds to the isotropic equivalent energy measured by an on-axis observer.}, $n$ is the circumburst density at a distance of $10^{17}$~cm, $p$ is the power-law index of the electron energy-distribution, $\theta_{1/2, \rm jet}$ is the jet half-opening angle, $\theta_{\rm obs}$ is the observer/viewing angle, $\epsilon_e$ and $\epsilon_B$ are the fractions of the internal energy in the shock-generated magnetic field and electrons, respectively, and $\xi_N$ is the fraction of electrons that are accelerated and $z$ is the redshift.

We fix the fractions of $\epsilon_e$ and $\xi_N$ at 0.1 and 1, respectively, and $p$ to 2.43 and the redshift to 0.009854. The other parameters are varied within the following ranges: $\theta_{1/2, \rm jet} = 5^{\circ}$--$45^{\circ}$, E$_{\rm AG,\,iso} = 10^{47}$--$10^{53}$~erg, $\epsilon_B = 10^{-5}$--$10^{-2}$, $n = 10^{-4}$--$10^{-1}$~cm$^{-3}$ and $\theta_{\rm obs} = 0^{\circ}$--$45^{\circ}$. The afterglow was modeled in a homogeneous ISM environment and we apply this model to eight representative frequencies: 1.4, 3, 6, 21, 338.5~GHz as well as the optical filters $F606W$ and $F475W$ and X-rays at 3~keV.

A critical aspect of the off-axis afterglow modeling is the resolution in azimuthal direction, in particular for models with large $\theta_{\rm obs}$ to $\theta_{1/2,\rm jet}$ ratios. We chose a numerical resolution of 20 and 30, for $\theta_{1/2,\,\rm jet}>9^\circ$ and $\theta_{1/2,\,\rm jet}<9^\circ$, respectively. Comparisons to simulations with a numerical resolution in azimuthal direction of 70 show that the lower resolution templates accurately capture the temporal evolution of the afterglow and they are also able to recover the absolute flux scale at maximum to within 20\%. The maximum flux of models with very narrow jet are recovered less accurately in off-axis afterglow models. As we show below, these models are not adequate to describe the observed afterglow evolution independent of the issue of the absolute flux scale. The numerical resolution in azimuthal direction was set to unity if the viewing angle is negligible, as suggested by the \package{boxfit} manual.

The gray curves in Fig. \ref{fig:lc} display a set of strict on-axis afterglow models (i.e., $\theta_{\rm obs}=0$) with a half-opening angle of $5^\circ$, $\epsilon_B=0.01$, $n=10^{-2}~{\rm cm}^{-3}$, and for $E_{\rm AG,\,iso}$ between $10^{48}$ and $10^{53}$~erg. Common to on-axis afterglow models ($\theta_{\rm obs}<\theta_{1/2,\,\rm jet}$) is the strict monotonic decline in X-rays and the optical, whereas the radio can exhibit a plateau or an initial rise. This evolution is in stark contrast to observations of AT2017gfo. The best-matching templates (colored curves in Fig. \ref{fig:lc}) strongly argue for a GRB seen off-axis (i.e., $\theta_{\rm obs}>\theta_{1/2,\,\rm jet}$; possible off-axis LGRB candidates were discussed in \citealt{Fynbo2004,Guidorzi2009a, Kruhler2009a}).

\textit{Model 1} (which we call the \textit{wide-jet model}, thin curves in Fig. \ref{fig:lc}) represents an afterglow with an energy reservoir of $\sim10^{50}~\rm erg$, an energy fraction stored in magnetic fields of $\epsilon_B\sim10^{-2}$, and a moderately collimated outflow with a half-opening angle of $\theta_{1/2,\,\rm jet}\sim20^\circ$, traversing a circumburst medium with a density of $10^{-2}~{\rm cm}^{-3}$. The jet axis and the line of sight are misaligned by $41^\circ$. \textit{Model 2} (which we call the \textit{narrow-jet model}, thick curves in Fig. \ref{fig:lc}) represents a more collimated jet with $\theta_{1/2,\,\rm jet}\sim5^\circ$, $E_{\rm AG,\, iso}\sim10^{51}~\rm erg$ and $\epsilon_B\sim2\times10^{-3}$,
traversing a more tenuous circumburst medium $(n\sim5\times10^{-4}~{\rm cm}^{-3})$. In this scenario, the line of sight and the GRB jet axis are misaligned by $17^\circ$.

The inferred afterglow properties are in both cases very close to the average values of SGRBs in \citet{Fong2015}, corroborating that this GRB is \textit{not} different from the population of known SGRBs. The properties of the wide-jet model are consistent with \citet{Alexander2017b}, \citet{Granot2017a}, \citet{Margutti2017b} and \citet{Troja2017a}. However, the two distinct models, discussed in this paper, show that there is significant degeneracy between the afterglow parameters \citep[for a more detailed study of the afterglow parameter space see][]{Granot2017a}. More detections are required to constrain the parameter space better. We note that the derived viewing angles for both models are consistent with the conservative limit of $<56^\circ$ from the LIGO signal. Moreover the narrow jet template (\textit{Model 2}) is consistent with the even stricter LIGO limit of $<28^\circ$ \citep{LIGO_VIRGO_1}.

With the best-match templates in hand, we quantify the contamination of the kilonova by the afterglow. The upper panels in Fig. \ref{fig:lc} display the light curve in $F475W$ ($6.2\times10^{12}$~Hz) and $r'$/$F606W$ ($4.9\times10^{12}$~Hz) by \citet{Tanvir2017b}. The contamination by the afterglow in the optical is negligible ($<1\%$) during the week after GW170817 for both models. Hence the inferred KN properties in \citet{Tanvir2017b} do not require any afterglow correction.

\begin{figure*}
\begin{center}
\includegraphics[width=1\columnwidth]{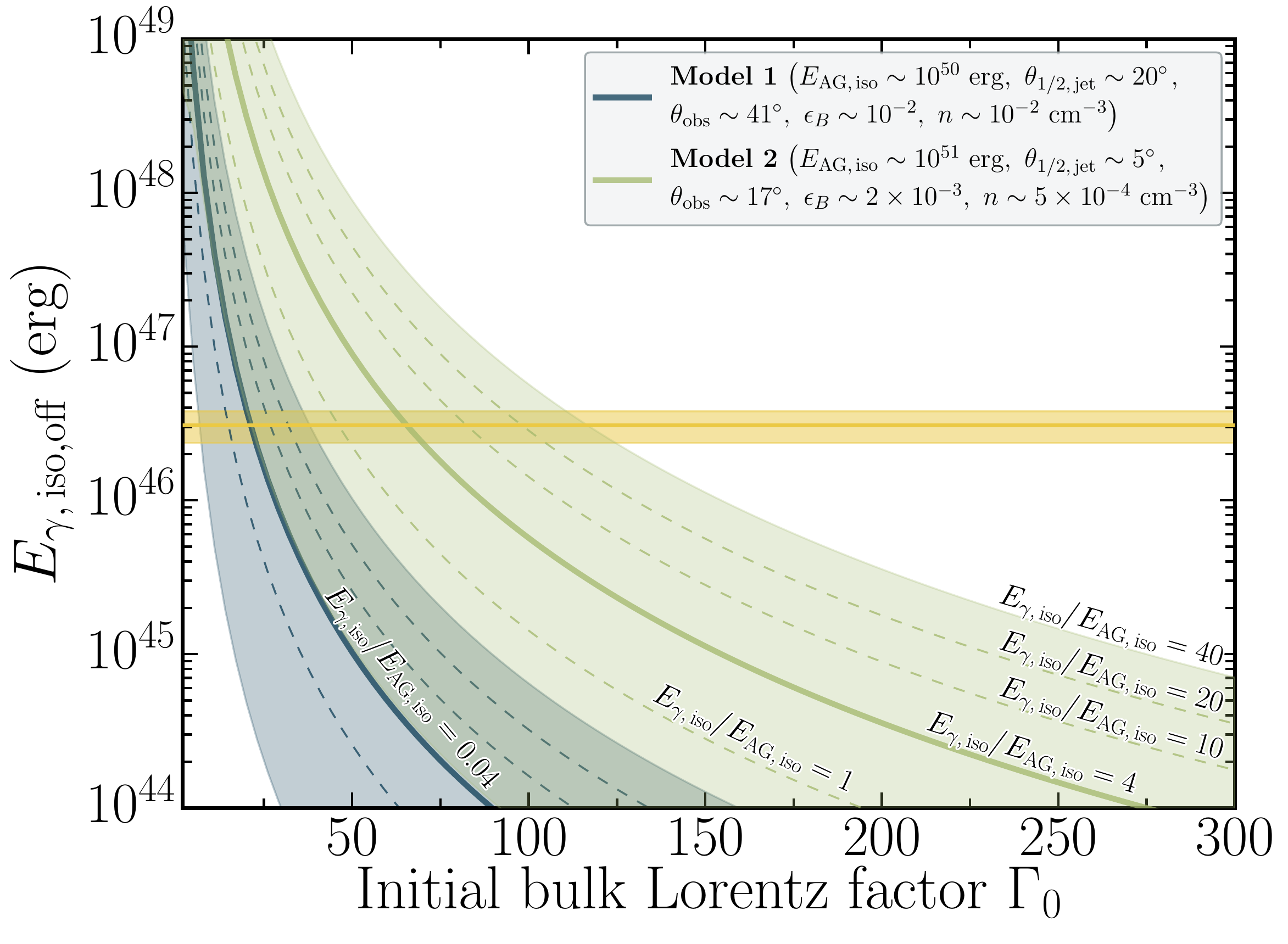}
\includegraphics[width=1\columnwidth]{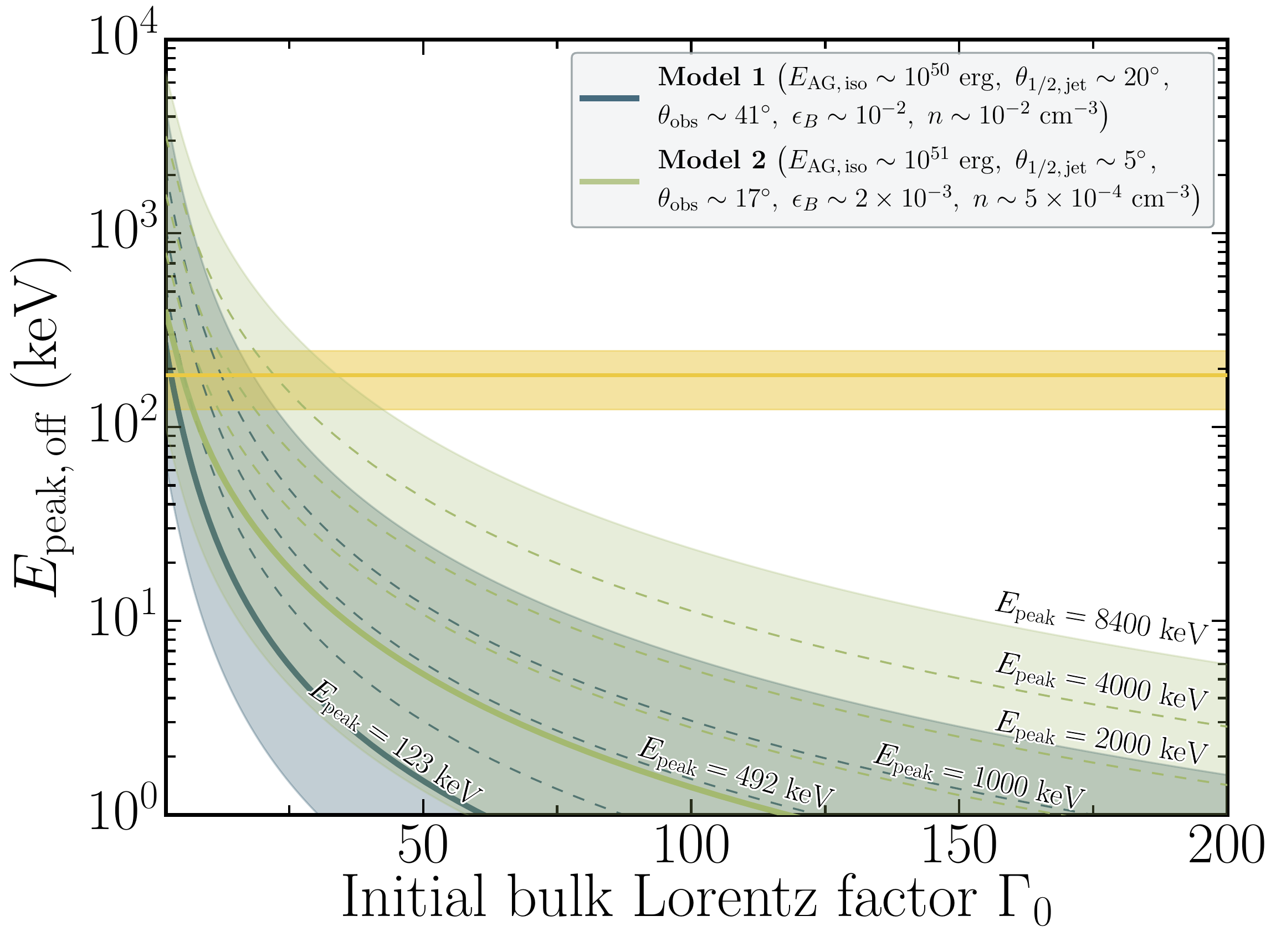}
\caption{The isotropic $\gamma$-ray energy release $E_{\gamma \rm{iso, off}}$ and the $E_{\rm peak}$ of the prompt emission measured by an off-axis observer, as a function of $\Gamma_0$. We use the model in \citet{donaghy05} to predict the allowed loci of the parameter spaces for GRB 170817A. The loci depend on the jet geometry and the viewing angle which we obtained from the broadband modeling in $\S$\ref{sec:modeling}. The known distributions for $E_{\gamma,\,\rm iso}/E_{\rm AG,\,iso}$ \citep{cenko2011} as well as the observed peak energy distribution of our SGRB comparison sample further limit the possible parameter space to the shaded regions (\textit{Model 1}: blue; \textit{Model 2}: red). Curves of particular values are displayed by the dashed curves. The solid curves indicates the location of the distribution mean values. The parameter space is furthermore limited by the results from \fermi\ reported in \citet{Goldstein2017b}, displayed in yellow. The bands encompassing the \fermi\ measurements indicate the $1\sigma$ confidence intervals. These constraints limit $\Gamma_0$ to $<15$ and $<43$ for \textit{Model 1} and \textit{Model 2}, respectively.}
\label{fig:prompt}
\end{center}
\end{figure*}

\subsection{Inferring jet parameters from $\gamma$-ray and afterglow emission}

The observed energy release of a GRB, $E_{\gamma,\,\rm iso,\,off}$, measured by an observer at a viewing angle $\theta_{\rm obs}$, depends on $\theta_{\rm obs}$, $\theta_{1/2, \rm jet}$, the jet geometry, and the initial bulk Lorentz factor of the jet, $\Gamma_0$. Similarly, the observed $E_{\rm peak,\,off}$ is a function of the same quantities. The simplest jet model assumes a uniform jet with a negligible surface and predicts that the ratios between $E_{\gamma,\,\rm iso,off}$ and on-axis $E_{\gamma,\,\rm iso}$, as well as $E_{\rm peak, off}$ and on-axis $E_{\rm peak}$ are simple powers of the Doppler factor of the jet \citep{Troja:nature,GRB2017a}. However, results already start to change considerably if the finite size of the jet is taken into account \citep{Yamazaki02,Yamazaki03}.

In the following, we assume a top-hat jet, similar to  \citet{Troja:nature} and \citet{GRB2017a}, but we take  into account the finite size of the jet. \citet{donaghy05} and \citet{graziani05} provided analytical expressions for $E_{\gamma,\,\rm iso,\,off}$ and $E_{\rm peak, off}$ for such a geometry that also match well the numerical calculations in \citet{Yamazaki02,Yamazaki03}. In this model, $E_{\gamma,\,\rm iso,\,off}$ and $E_{\rm peak,\,off}$ are given by

\begin{eqnarray}
E_{\gamma,\,\rm iso,\,off} &=& \frac{E_{\gamma,\,\rm iso}}{2\beta\Gamma_0^{4}}
	\left[ f(\beta-\cos\thetav) - f(\beta\cos\thetan-\cos\thetav) 
	\right], \nonumber\\
\epobs &=& \frac{\left\langle D (\thetav)\right\rangle}{\left\langle D (\thetav=0)\right\rangle} \ep, \nonumber
\end{eqnarray} 
where $\beta$ the velocity normalized to the speed of light, $\Gamma$ is the Lorentz factor, the function $f(y)$ is defined as
\begin{equation}
f(y) = \frac{\Gamma_0^{2}(2\Gamma_0^{2}-1)y^{3} + (3\Gamma_0^{2}\sin^{2}\thetav)y
	+ 2\cos\thetav\sin^{2}\thetav}{(y^{2} +
	\Gamma_0^{-2}\sin^{2}\thetav)^{3/2}},\nonumber
\end{equation}
and the average Doppler shift is given by
\begin{equation}
\davg = \Gamma_0^{-1} 
	\frac{f(\beta-\cos\thetav) - f(\beta\cos\thetan-\cos\thetav)}
	{g(\beta-\cos\thetav) - g(\beta\cos\thetan-\cos\thetav)},\nonumber
\end{equation}
with
\begin{equation}
g(y) = \frac{2\Gamma^{2}y + 2\cos\thetav}
	{(y^{2} + \Gamma^{-2}\sin^{2}\thetav)^{1/2}}\nonumber.
\end{equation}

The equations for $E_{\gamma,\,\rm iso,\,off}$ and 
$\epobs$ depend on the unknown $E_{\gamma,\,\rm iso}$, $\ep$ and the initial bulk Lorentz factor $\Gamma_0$. Considering the complexity of the expressions, we perform a parameter study. We can limit the possible parameter space by using results from other SGRBs and our afterglow modeling. \citet{cenko2011} reported that the ratio between $E_{\gamma,\,\rm iso}$ and $E_{\rm AG,\, iso}$ varies between 0.05 and 40 (mean value being $\sim4$). A $E_{\gamma,\,\rm iso}$-$E_{\rm AG,\, iso}$ ratio of a few has also been observed for SGRBs \citep{Fong2015}. The broadband modeling ($\S$\ref{sec:modeling}) suggest that $E_{\rm AG,\, iso}$ is between $10^{50}$~erg and $10^{51}$~erg, corresponding to $E_{\gamma,\,\rm iso}=5\times10^{48}-40\times10^{51}$~erg. The $\ep$ distribution of our SGRB comparison sample extends from $\sim40$ to $\sim8400$~keV (mean peak energy being $\sim490$~keV). \citet{Goldstein2017b} reported a peak energy of $185\pm62$~keV for the main emission of GRB 170817A. Hence, we vary $\ep$ between 123~keV and 8400~keV.

In the left and right panels of Fig. \ref{fig:prompt}, we display the $E_{\gamma,\,\rm iso,\,off}$ and $E_{\rm peak,\,off}$ as a function of $\Gamma_0$. The expected parameter spaces for the two afterglow models are shown in blue (wide-jet) and red (narrow-jet). Overlaid in yellow are the observed $E_{\gamma,\,\rm iso,\,off}
/E_{\rm peak,\, iso,\,off}$ values reported by \citet{Goldstein2017b}. The overlapping regions show the allowed parameter space for the GRB, if seen on-axis, for each afterglow model.

The observed span in the $E_{\gamma,\,\rm iso}$ and $E_{\rm AG,\, iso}$ allows initial bulk Lorentz factor between 6 and 40, and 20 and 125 for the wide-jet (\textit{Model 1}) and narrow-jet (\textit{Model 2}), respectively (left panel in Fig. \ref{fig:prompt}). However, the highest Lorentz factor would always require $E_{\gamma,\,\rm iso}/E_{\rm AG,\, iso}\gtrsim10$. While such values are not atypical they are at the upper end of the observed distribution in \citet{Fong2015}. The observed distribution of peak energies of short GRBs narrows the possible parameter space further: $\Gamma_0<15$ and $\Gamma_0<43$ for \textit{Model 1} and \textit{Model 2}, respectively (right panel in Fig. \ref{fig:prompt}), while the required peak energies need to exceed at least several hundred keV. The Lorentz factors are similar to the values in \citet{zou2017} who assumed a top-jet with negligible surface and used the $E_{\rm peak}$-$E_{\gamma,\,\rm iso}$ (derived from SGRBs) and $\Gamma_0$-$E_{\gamma,\,\rm iso}$ (derived from LGRBs) correlations.

To understand these results, we reflect upon the assumptions of this calculation. This parameter study of the jet parameters depends on the jet half-opening angle, the viewing angle and the jet geometry. The model in \citet{donaghy05} and \citet{graziani05} assumes that the $\gamma$-ray emission is produced via internal shocks in the GRB jet. According to \citet{graziani05}, systematic uncertainties may exist between the $\epobs$ calculated by the above expression and the observed peak of the effective GRB spectrum. Therefore it may not always be a very accurate representation of the observed $\ep$. Furthermore, the parameter space is limited to the observed $E_{\rm peak}$ distribution of short GRBs and the known ratios between $E_{\gamma,\,\rm iso}$ and $E_{\rm AG,\, iso}$ and the parameters of GRB170817 as measured by \fermi.

Our broadband modeling is based on a small number of detections and we showed that there is substantial degeneracy in the model parameters. This degeneracy is also visible in the $\Gamma_0$-$E_{\gamma,\,\rm iso,\,off}$  and the $\Gamma_0$-$E_{\rm peak,\,off}$ parameter spaces (Fig. \ref{fig:prompt}). A more sophisticated afterglow modeling and the inclusion of more afterglow observations can reduce the degeneracy. The conclusions of this analysis also depend significantly on the observed $\gamma$-ray properties of GRB170817A. \citet{Goldstein2017b} reported an error of 33\% on the peak energy. A substantially lower peak energy would allow higher $\Gamma_0$ for lower peak energies \citep[For an independent analysis of the \fermi/GBM data see][]{zhang2017}.

\subsection{Low frequency radio emission from merger ejecta}
Non-relativistic shocks from the merger ejecta are thought to emit at radio frequencies \citep{nakar2011}. This model predicts that the emission peaks in the MHz regime and at the epoch of deceleration of the non-relativistic shock, which is expected to be on the order of months to years after the merger.

To examine whether this mechanism could produce a bright transient months after GW170817, we use the observed properties of the kilonova and the afterglow. The expected brightness at optically thin GHz frequencies would be,
\begin{equation}
f_{\nu} (t) = 655{\rm mJy} \; {n_0}^{1.8} {\frac{R(t)}{10^{17}}}^{3}  {\beta(t)}^{3.75} \epsilon_B^{0.78} \epsilon_e \, \left(\frac{\nu}{\rm GHz} \right)^{-0.55}, \nonumber
\end{equation}
for an electron index of 2.1, where $R(t)$ is the radius of the shock front (normalized to $10^{17}$~cm) and $\beta (t)$ is the velocity normalized to the speed of light $c$. This expression is derived from the peak synchrotron flux and the characteristic synchrotron frequency $\nu_m$ of the power-law electron distribution, and is in agreement with expressions in \citet{nakar2011}.
 The radius and the observed time $t$ are related through $R(t) = \beta c t$. The epoch of deceleration,  where the swept-up mass equals the ejected mass, is given by $t_{\rm dec} = 7\, {\rm yr} \left( \frac{M_{\rm ej, \odot}}{n_0} \right)^{1/3} \beta_0^{-1}$, where $\beta_0$ is the normalized initial velocity.

\citet{Tanvir2017b} concluded that the merger ejected $\sim5\times10^{-4}~M_\odot$ with a velocity of $0.1\,c$. Along with a circumburst medium density, $n = 0.01~{\rm cm}^{-3}$, we estimate the brightness at $1.4$~GHz to be $\sim60~\mu\rm Jy$ for a deceleration time scale of 55 years. A smaller ambient density will further reduce the flux and increase the $t_{\rm dec}$. 

Considering the results of \citet{Smartt2017a}, where a higher ejected mass of $0.01M_{\odot}$ was estimated to be released with a similar $\beta$, the deceleration time will be $\sim 70$~yr, and the observed flux will remain the same as it is insensitive to $M_{\rm ej}$. Therefore, the outlook, assuming this model is valid, is bleak. 

The merger remnant, if a magnetar, can inject additional energy into the shock \citep{mb2014}. This increased energy will also delay  $t_{\rm dec}$. In this model, from the observed $\beta = 0.1$ and best-fit ambient density $n_0 = 0.01~{\rm cm}^{-3}$, $t_{\rm dec}=260\,E_{\rm mag, 52}$~yr, where $E_{\rm mag, 52}$ is the energy input from the magnetar. The peak flux in $1.4$~GHz at $t_{\rm dec} \sim 3$~mJy for our parameters, which like in the previous case scales down as a $t^3$ power-law to the current epoch.  Therefore we do not expect any detectable emission at GMRT frequencies at present from the merger ejecta, consistent with our observations. 

\section{Summary}

LIGO/Virgo detected a BNSM at a distance of $\sim44$~Mpc on 17 August 2017. Rapid optical and near-IR follow-up observations detected a new transient, AT2017gfo, in the credible region of GW170817 with properties consistent with KN models. Gamma-ray satellites detected the short GRB 170817A quasi-contemporaneously with GW170817, but owing to the poor localization at $\gamma$-rays this did not exclude a chance alignment.

We observed the position of AT2017gfo with ALMA and GMRT at 338.5 and 1.4 GHz, respectively, from 1.4 days to 44 days after the merger, our objective being to constrain the GRB afterglow component. The afterglow evaded detection at all epochs. Our radio and sub-mm observations allow us to place a firm upper limit of a few $10^{26}~{\rm erg\,s}^{-1}$ in the sub-mm and radio, probing a regime $>2$--4 orders of magnitudes fainter than previous limits on SGRBs.

The emergence of an X-ray and radio transient at the position of AT2017gfo at 9 and 17 days after GW170817, respectively, is highly atypical for GRBs. Modeling the evolution from radio to X-ray frequencies with templates generated from 2D relativistic hydrodynamical jet simulations excludes all on-axis afterglow models ($\theta_{1/2,\,{\rm jet}} > \theta_{\rm obs} $) with sensible physical parameters. Adequate models, describing the evolution from X-ray to radio frequencies, require strict off-axis afterglow templates where $\theta_{1/2,\,{\rm jet}} < \theta_{\rm obs}$. \textit{Model 1} favors a jet, powered by $E_{\rm AG,\,iso}\sim10^{50}~\rm erg$, with magnetic equipartition of $\epsilon_B\sim10^{-2}$ and an initial half-opening angle of $\sim20^\circ$, traversing a circumburst medium with $n=10^{-2}~{\rm cm}^{-3}$. The second model suggests a more collimated jet: $E_{\rm AG,\,iso}\sim10^{51}~\rm erg$, $\theta_{1/2,\, jet}\sim5^\circ$, $\epsilon_B=2\times10^{-3}$, $n=5\times10^{-4}~{\rm cm}^{-3}$. More detections of the afterglow are needed to reduce the degeneracy in the model parameters. In both cases our line of sight and the GRB jet axis were misaligned, by $\sim41^\circ$ (wide-jet model) and $\sim20^\circ$ (narrow-jet model), explaining the emergence of the afterglow only a week after the GRB. The viewing angle measurements are consistent with upper limits by \citet{LIGO_VIRGO_1}.

The jet parameters are, in both cases, consistent with mean values of the \swift\ SGRB population. Using $\theta_{1/2,\,{\rm jet}}, \theta_{\rm obs}, $ and $E_{\rm AG,\,iso}$ of the blast wave, we inferred the true $\gamma$-ray energy release and initial bulk Lorentz factor ($\Gamma_0$) of the flow. We find evidence for an ultra-relativistic jet with $\Gamma_0 <15/<43$ for \textit{Model1/2}. The prompt energy release has to be at least a factor of a few higher than the kinetic energy in the afterglow and peak energies of least several hundred keV that an on-axis observer would have recorded. Therefore, we conclude that a uniform top-hat jet model can broadly explain  the observed gamma-ray properties of GRB170817A. Limiting this parameter study is the degeneracy in the afterglow parameters, due to the limited amount of data, and the large uncertainties of the observed peak energy.

Using the best-match template we assessed if the afterglow contaminated significantly the KN optical emission. The contamination is $<1\%$ during the first week after GW170817. We also calculated the expected radio emission from the merger ejecta and found it to be negligible presently.

The afterglow modeling allows us to draw the following conclusions:
\textit{i}) this is the first robust detection of an off-axis GRB with $\theta_{\rm obs} > \theta_{1/2,\,\rm jet}$ and \textit{ii}) AT2017gfo and GRB~170817A have the same progenitor.
These findings in conjunction with the spectroscopic evidence for $r$-process elements in spectra of AT2017gfo \citep[e.g.,][]{Pian2017a, Smartt2017a}, demonstrate that some SGRBs are connected with BNSMs and firmly establishes that AT2017gfo and GRB~170817A are the electromagnetic counterpart to GW170817.

\acknowledgments
We thank the referee for a careful reading of the manuscript
and for helpful comments that improved this paper.

We gratefully acknowledge support from:
FONDECYT grant 3130488 (SK),
FONDECYT grant 3160439 (JB),
CONICYT grants Basal-CATA PFB-06/2007 (SK, FEB, JB), 
FONDECYT Regular 1141218 (FEB, JG-L), and Programa de Astronomia FONDO ALMA 2016 31160033 (JG-L); 
the Ministry of Economy, Development, and Tourism's Millennium Science Initiative through grant IC120009, awarded to The Millennium Institute of Astrophysics, MAS (FEB, JB);
the Science and Technology Facilities Council (ABH, RLCS);
the Spanish research project AYA 2014-58381-P (CCT, AdUP, DAK); 
the Ram\'on y Cajal fellowship RyC-2012-09975 (AdUP);
the Ram\'on y Cajal fellowship RyC-2012-09984 (CCT);
the 2016 BBVA Foundation Grant for Researchers and Cultural Creators (AdUP);
the Juan de la Cierva Incorporaci\'on fellowship IJCI-2015-26153 (DAK);
a VILLUM FONDEN Investigator grant (project number 16599) (JH);
the National Science Centre, Poland through the POLONEZ grant 2015/19/P/ST9/04010 (MJM);
the European Union's Horizon 2020 research and innovation programme under the Marie Sk{\l}odowska-Curie grant agreement No. 665778 (MJM);
the Leverhulme Trust Early Career Fellowship (SRO);
SC acknowledge partial funding from ASI-INAF grant I/004/11/3;
RS-R acknowledges support from ASI (Italian Space Agency) through the Contract n. 2015-046-R.0 and from European Union Horizon 2020 Programme under the AHEAD project (grant agreement n. 654215).

ALMA is a partnership of ESO (representing its member states), NSF (USA) and NINS (Japan), together with NRC (Canada), MOST and ASIAA (Taiwan), and KASI (Republic of Korea), in cooperation with the Republic of Chile. The Joint ALMA Observatory is operated by ESO, AUI/NRAO and NAOJ.


The GMRT is run by the National Center for Radio Astrophysics of the Tata Institute of Fundamental Research.

L. Resmi thanks Suma Murthy and Swagat R. Das for discussions on radio interferometric analysis.

This work made use of data supplied by the UK \swift\ Science Data Centre at the University of Leicester.

Development of the Boxfit code was supported in part by NASA through grant NNX10AF62G issued through the Astrophysics Theory Program and by the NSF through grant AST-1009863. Simulations for BOXFIT version 2 have been carried out in part on the computing facilities of the Computational Center for Particle and Astrophysics (C2PAP) of the research cooperation "Excellence Cluster Universe" in Garching, Germany.

\facilities{ALMA, GMRT} 
\software{AIPS, Boxfit, CASA, MultiNest, PyMultiNest}

\bibliography{mybib}

\end{document}